\documentclass[preprint,3p,12pt,authoryear]{elsarticle}

\usepackage{amssymb}
\usepackage{amsmath}
\usepackage{multirow}
\usepackage{xcolor} 
\usepackage{url}

\newtheorem{theorem}{Theorem}
\newtheorem{lemma}{Lemma}
\newproof{proof}{Proof}

\newcommand{\R}{\mathbb R}
\newcommand{\Sbb}{\mathbb S}

\newcommand{\C}{\mathcal C}

\newcommand{\ab}{\mathbf a}
\newcommand{\bb}{\mathbf b}
\newcommand{\cb}{\mathbf c}
\newcommand{\db}{\mathbf d}
\newcommand{\e}{\mathbf e}
\newcommand{\f}{\mathbf f}

\newcommand{\h}{\mathbf h}

\newcommand{\w}{\mathbf w}
\newcommand{\x}{\mathbf x}
\newcommand{\y}{\mathbf y}
\newcommand{\z}{\mathbf z}

\newcommand{\A}{\mathbf A}
\newcommand{\B}{\mathbf B}

\newcommand{\E}{\mathbf E}
\newcommand{\F}{\mathbf F}
\newcommand{\G}{\mathbf G}
\newcommand{\Hb}{\mathbf H}
\newcommand{\I}{\mathbf I}

\newcommand{\Lb}{\mathbf L}
\newcommand{\M}{\mathbf M}
\newcommand{\N}{\mathbf N}

\newcommand{\Q}{\mathbf Q}

\newcommand{\U}{\mathbf U}

\newcommand{\X}{\mathbf X}
\newcommand{\Y}{\mathbf Y}
\newcommand{\Z}{\mathbf Z}

\newcommand{\Sigmab}{\boldsymbol \Sigma}
\newcommand{\betab}{\boldsymbol \beta}

\newcommand{\Bf}{\mathfrak B}

\newcommand{\0}{\mathbf 0}
\newcommand{\1}{\mathbf 1}

\newcommand{\tr}{\mathrm{tr}}

\newcommand{\vv}{\mathrm{vec}}
\newcommand{\vech}{\mathrm{vech}}

\journal{Journal of Statistical Planning and Inference}

\begin{document}

\begin{frontmatter}

\title{Mixed-integer linear programming for computing optimal experimental designs}

\author[label1]{Radoslav Harman}
\ead{radoslav.harman@fmph.uniba.sk}

\author[label1]{Samuel Rosa\corref{cor1}}
\ead{samuel.rosa@fmph.uniba.sk}

\cortext[cor1]{Corresponding author}

\affiliation[label1]{organization={Comenius University Bratislava},
	city={Bratislava},
	country={Slovakia}}

\begin{abstract}

The problem of computing an exact experimental design that is optimal for the least-squares estimation of the parameters of a regression model is considered. We show that this problem can be solved via mixed-integer linear programming (MILP) for a wide class of optimality criteria, including the criteria of A-, I-, G- and MV-optimality. This approach improves upon the current state-of-the-art mathematical programming formulation, which uses mixed-integer second-order cone programming. The key idea underlying the MILP formulation is McCormick relaxation, which critically depends on finite interval bounds for the elements of the covariance matrix of the least-squares estimator corresponding to an optimal exact design. We provide both analytic and algorithmic methods for constructing these bounds. We also demonstrate the unique advantages of the MILP approach, such as the possibility of incorporating multiple design constraints into the optimization problem, including constraints on the variances and covariances of the least-squares estimator.

\end{abstract}


\end{frontmatter}

\section{Introduction}

The optimal design of experiments is an important part of theoretical and applied statistics (e.g., \cite{Fedorov}, \cite{Pazman86}, \cite{puk}, \cite{AtkinsonEA07}, \cite{GoosJones}) that overlaps with various areas of general optimization (\cite{VanderbergheBoyd}, \cite{GhoshEA08}, \cite{Todd16}, and others). From the statistical perspective, the typical problem in optimal experimental design is the selection of trials to maximize the information obtained about the parameters of an underlying model.

In this research, we focus on optimal \textit{exact} designs, which are technically solutions to a special class of discrete optimization problems;
in contrast, \textit{approximate} designs are continuous relaxations of exact designs. Precise formulations of both design classes are provided in Section \ref{ssOD}. In the remainder of this paper, we use the term ``design'' specifically to refer to an exact design. 

Analytical forms of optimal designs are difficult to derive; they have only been found for some special cases (e.g., \cite{Gaffke86}, \cite{NeubauerEA}, \cite{Bailey07}). Therefore, much attention is given to numerical computations. There are two general classes of algorithms for solving optimal design problems: heuristics and enumeration methods.

Heuristics aim to rapidly generate a reasonably efficient design but often fail to converge to an optimal one. Some of the most popular heuristics include various exchange methods (see, e.g., Chapter 12 in \cite{AtkinsonEA07} for a review of classical algorithms or \cite{HuangEA} for more recent developments) and methods inspired by physics or the natural world (\cite{Haines87}, \cite{DevonLin}, \cite{ChenEA22} and others). Another approach is based on a quadratic approximation of the criterion (\cite{HarmanFilova}, \cite{FilovaHarman20}).

A separate approach consists of first computing optimal approximate designs and subsequently rounding them. The advantage of this approach is that the optimal approximate design problem is continuous and typically convex. From the perspective of mathematical programming, this means that this problem can be handled through various powerful methods of convex optimization, such as general convex programming (e.g., \cite{WongZhou}, \cite{WongZhou23}), semidefinite programming (e.g., \cite{VanderbergheBoyd}), second-order cone programming (\cite{Sagnol}, \cite{SagnolHarman15}) and linear programming (e.g., \cite{HarmanJurik}). The disadvantage is that for practical applications, the resulting optimal approximate designs need to be converted into exact designs via a rounding procedure (e.g., \cite{PukelsheimRieder92}), which often leads to a suboptimal exact design. That is, the two-step approach---first computing optimal approximate designs and then rounding them to exact designs---can also be considered a heuristic.

Compared to heuristics, enumeration methods are typically slower, but they ultimately find optimal designs with a proof of optimality. Enumeration methods include algorithms for solving integer programs because optimal design problems are, trivially, special instances of a general integer program. As such, very small optimal design problems can be solved through the complete enumeration of all permissible designs (e.g., \cite{HainesClark}), but a more economical integer or mixed-integer programming method is usually needed, such as branch and bound (e.g., \cite{Welch}, \cite{DuarteEA}, \cite{Ahipasaoglu21}). For problems that are shown to have a special structure, specialized and, thus, typically better-performing solvers can be employed. In particular, a mixed-integer second-order cone programming (MISOCP) formulation of the optimal design problem for the most popular design criteria presented by \cite{SagnolHarman15} permits the use of efficient branch and cut algorithms.
\bigskip

In this paper, we show that a large class of optimal design problems can be solved by means of an even more specialized mixed-integer \textit{linear} programming (MILP) approach. More precisely, we show that the classical exact optimal design problem can be expressed as an MILP problem for a rich class of minimax optimality criteria, including the fundamental criteria of $A$-, $I$-, $MV$- and $G$-optimality. 

This result is theoretically important, but it also has practical advantages. Most obviously, in contrast to MISOCP solvers, MILP solvers are more readily available. Moreover, our MILP formulation allows one to introduce additional constraints on the designs corresponding to practical requirements---these could be not only linear constraints on the experimental resources, but also limits on the variances and covariances of the least-squares estimator (more details are provided in Section \ref{sec:Extend}). In contrast, heuristics cannot handle such additional design constraints (with some exceptions; see, e.g., \cite{HarmanEA16}), and even the existing enumeration methods (such as those of \cite{Ahipasaoglu21} and \cite{SagnolHarman15}) typically do not directly allow such a wide range of constraints (and it would be nontrivial to modify them to do so).
Another advantage of the MILP formulation is that it covers criteria such as $MV$-optimality, for which an MISOCP formulation is currently unavailable.

Integer linear programming (ILP) and MILP have already been used in the field of experimental design for various purposes, albeit in a substantially different way than the one proposed in this paper. For instance, ILP has been used for the classification of orthogonal arrays and for the generation of minimum-size orthogonal fractional factorial designs (e.g., \cite{BulutogluMargot}, \cite{Fontana13}), and MILP has been used for optimal arrangements of orthogonal designs (e.g., \cite{SartonoEA15}, \cite{SartonoEA152}, \cite{VoThanhEA}). However, the above applications utilize specific (linear) characteristics of the studied models or of the studied classes of designs, whereas we provide an MILP formulation for arbitrary models, i.e., for problems that are not inherently linear (although we utilize certain indirect linear aspects of the considered optimality criteria, as detailed in Section \ref{ssOD}).
\bigskip

The key idea underlying our formulation is McCormick relaxation (\cite{McCormick}), which critically depends on finite interval bounds for all elements of a certain matrix. In the context of the optimal design of experiments, this matrix is the covariance matrix of the least-squares estimator of the model parameters under the optimal design; for brevity, we call such constraints \textbf{covariance bounds}.
It would be trivial to construct covariance bounds for the MILP problem provided that we knew the optimal design. However, the optimal design is the final aim of the MILP computation itself; therefore, it is not known beforehand. The situation is also complicated by the fact that the covariance matrix depends on the experimental design in a nonlinear way. Therefore, the construction of covariance bounds is a major challenge in the application of the MILP approach, and we provide both analytical and numerical constructions of such bounds. 

MILP formulations based on McCormick relaxation were recently proposed by \cite{BhelaEA} in the network topology context and, independently, by \cite{OkasakiEA} in the context of $c$-optimal Bayesian sampling designs. \cite{BhelaEA} sought an optimal placement of edges in a graph so as to minimize a given function of the eigenvalues of the (weighted) Laplacian. This problem corresponds to $A$-optimality in a specific statistical model (for more details on the connection between optimal designs and optimizing graph Laplacians, see \cite{Cheng81} and \cite{BaileyCameron}). Thus, the problems addressed by \cite{BhelaEA} and \cite{OkasakiEA} can mathematically be viewed as special instances of the optimal design problem.
In this paper, we extend their approaches to general problems of experimental design and to a broad class of minimax criteria. Importantly, to this end, we provide the covariance bounds for such general problems.
\bigskip

In the remainder of this section, we introduce the notation used in this paper and discuss the optimal design problem in more detail. In Section \ref{sMILP}, we derive the MILP formulation, and in Section \ref{sBounds}, we provide covariance bounds that can be used for this formulation. Various extensions are discussed in Section \ref{sec:Extend}. The proposed formulation is applied in multiple experimental settings and compared with the MISOCP approach in Section \ref{sStudy}. Finally, Section \ref{sDiscussion} presents a short discussion.

\subsection{Notation}

Throughout the paper, we use the following standard mathematical notation: $\I_n$ is the $n \times n$ identity matrix; $\0_{m \times n}$ is the $m \times n$ matrix of zeros; $\1_n$ is the $n \times 1$ all-ones vector; and $\0_n$ is the $n \times 1$ vector of zeros. Indices are suppressed where they are evident. The symbol $\e_i$ denotes the $i$-th elementary unit vector (i.e., its $i$-th element is one, and all other elements are zero); $\mathbb{R}^n$ is the set of all real $n \times 1$ vectors; and $\Sbb^m$, $\Sbb^m_+$ and $\Sbb^m_{++}$ are the sets of all $m \times m$ symmetric, (symmetric) nonnegative definite and (symmetric) positive definite matrices, respectively. For $\M \in \Sbb^m_+$, $\lambda_{\max}(\M)$ denotes the maximum eigenvalue, $\tr(\M)$ is the trace, $\M^{1/2}$ is the square-root matrix, and if $\M$ is nonsingular, $\M^{-1/2}$ denotes the inverse of $\M^{1/2}$. For any $m \times n$ matrix $\A$, we denote its transpose by $\A'$ and its Moore--Penrose pseudoinverse by $\A^+$; $\C(\A)$ is the column space (the range) of $\A$. The notation $(\A)_{jk}$ represents the element of $\A$ with coordinates $(j,k)$. The symbol $\|\cdot\|$ denotes the Euclidean norm of a vector or the Frobenius norm of a matrix. The symbol $\mathbb{N}_0$ represents the set of all nonnegative integers, and $\{1{:}n\}$ is an abbreviated form of the set $\{1,\ldots,n\}$.

\subsection{Optimal designs}
\label{ssOD}

For clarity, we explain the main message of this paper in the context of the standard optimal experimental design on a finite design space $\{1{:}n\}$. That is, suppose that under the $i$-th experimental conditions (``at design point $i$''), the real-valued observation satisfies the regression model $y_i = \f_i'\betab + \varepsilon_i$, where $\betab \in \R^m$ is the vector of unknown parameters and the $\f_i \in \mathbb{R}^m$, $i \in \{1{:}n\}$, are known vectors, sometimes called regressors. An actual experiment then consists of $N$ trials performed at $N$ design points selected from $\{1{:}n\}$.
We assume that $m \geq 2$, $E(\varepsilon_i)=0$, $D(\varepsilon_i)=1$ and that the observations from distinct trials are uncorrelated. The assumption $D(\varepsilon_i)=1$ can be made without loss of generality and is therefore commonly adopted in the optimal design of experiments. In Section \ref{sec:Extend}, we show that our results can be easily extended to more general settings, such as nonlinear models, continuous design spaces and diverse design constraints. We note, however, that they do not extend to the case of correlated observations because for such problems, the information matrix does not have a desirable (additive) form.

We also initially restrict ourselves to binary (i.e., replication-free) exact designs, as the corresponding optimal design problems naturally lend themselves to our MILP reformulation. Optimal binary designs are of interest by themselves (see, e.g., \cite{RaschEA}); however, we show in Section \ref{sec:Extend} that our results also extend to exact designs with replications. A binary design of size $N \leq n$ on $\{1{:}n\}$ is a selection of $N$ distinct elements of $\{1{:}n\}$. This can be formalized by representing a binary design as a vector $\db = (d_1,\ldots,d_n)' \in \{0,1\}^n$ such that $\1'_n\db=N$, where $d_i$ denotes the number of trials at design point $i$. The set of all such designs is denoted by $\mathcal{D}^{(n)}_N$. In contrast, a general design $\db$ satisfies  $\db \in \{0, 1, \ldots, N\}^n$ such that $\1'_n\db=N$. As we focus on binary designs in our presentation, by a ``design'' we mean a binary design.

The amount of information provided by a design $\db$ is measured by its information matrix $\M(\db) = \sum_{i=1}^n d_i \f_i \f_i'$, which, if nonsingular, is the inverse of the covariance matrix $\Sigmab$ of the least-squares estimator $\widehat{\betab}$ of $\betab$. Moreover, if the errors are normal, $\M(\db)$ is exactly the Fisher information matrix for $\betab$. We also suppose that $m \leq N$ and that the linear span of $\{\f_i: i \in \{1{:}n\}\}$ is the full $\mathbb{R}^m$, which guarantees that there exists at least one $\db \in \mathcal{D}^{(n)}_N$ with a nonsingular information matrix.

In some of our results, approximate designs will also play a minor role. 
Instead of integer numbers of trials, such designs represent the proportions of trials at individual design points for $N \to \infty$. An approximate design for a model $y_i = \f_i'\betab + \varepsilon_i$, $i \in \{1{:}n\}$, can be formally represented by a vector $\w \in \R^n$ such that $w_i \geq 0$ for all $i$ and $\sum_i w_i = 1$. The information matrix of $\w$ is then $\M(\w) = \sum_i w_i \f_i\f_i'$. We will always use the qualifier ``approximate'' when referring to an approximate design.

The ``loss'' associated with an information matrix is measured by a so-called optimality criterion $\Phi$, selected in accordance with the aim of the experiment; in our case, the optimality criterion is formally a function mapping from $\Sbb^m_{++}$ to $\R$. A $\Phi$-optimal design $\db^*$ then minimizes $\Phi(\M(\db))$ over all designs $\db$ of size $N$ with a nonsingular information matrix $\M(\db)$. Examples include $D$-optimality, for which $\Phi_D(\M) = \ln(\det(\M^{-1}))$; $A$-optimality, for which $\Phi_A(\M) = \tr(\M^{-1})$; $I$-optimality,\footnote{$I$-optimality is sometimes also called $V$-optimality or $IV$-optimality because it can be interpreted as an integrated prediction variance.}
for which $\Phi_I(\M) = \sum_{i=1}^n \f_i' \M^{-1}\f_i$; $G$-optimality, for which $\Phi_G(\M) = \max_{i=1,\ldots,n} \f_i' \M^{-1} \f_i$; $MV$-optimality, for which $\Phi_{MV}(\M) = \max_{i = 1,\ldots,m} \e_i' \M^{-1} \e_i$; $E$-optimality, for which $\Phi_E(\M)=\lambda_{\max}(\M^{-1})$; and $c$-optimality, for which $\Phi_c(\M) = \cb' \M^{-1} \cb$ for a selected $\cb \in \R^m$. For statistical interpretations of these criteria, see, e.g., Chapter 10 in \cite{AtkinsonEA07} and Chapter 5 in \cite{PronzatoPazman}.
\bigskip

In this paper, we focus on a broad class of criteria, which we now define. For $\ell \in \{1{:}K\}$, let $\B_\ell$ be a given $m \times s_\ell$ matrix. We use $\Bf$ to denote the sequence $\B_1,\ldots,\B_K$ and $\B$ to denote the $m \times \sum_\ell s_\ell$ matrix $(\B_1, \ldots, \B_K)$. We assume that $\C(\B) = \R^m$ and that no column of $\B$ is equal to $\0_m$. We consider the following criterion:
\begin{equation*}
	\Phi_{\mathfrak{B}}(\M) = \max_{\ell=1, \ldots, K} \tr(\B_\ell' \M^{-1} \B_\ell),
\end{equation*}
which belongs to the family of minimax criteria discussed by \cite{Wong92}. The expression $\B_\ell' \M^{-1} \B_\ell$ represents the variance matrix of the least-squares estimator of $\B_\ell'\betab$, which implies that $\Phi_\Bf$ seeks to minimize the maximum of the $A$-optimality criterion values over all linear parameter systems $\B_\ell'\betab$, $\ell \in \{1{:}K\}$. Clearly, all the $\B_\ell'\betab$s are estimable under a design $\db$ if and only if $\M(\db)$ is nonsingular. As special cases of $\Phi_{\Bf}$, we can obtain the criteria of $A$-optimality ($K=1$, $\B_1 = \I_m$), $I$-optimality\footnote{The $I$-optimal design can also be found by computing the $A$-optimal design for a transformed model (see, e.g., Appendix A.3 in \cite{SagnolHarman15}). In particular, the $A$-optimal design for a model with regressors $\tilde{\f}_i = \G^{-1/2} \f_i$, $i \in \{1{:} n\}$, where $\G = \sum_{i=1}^n \f_i \f_i'$, is $I$-optimal for the original model. However, in this paper, we consider $I$-optimality as a standalone criterion. Note that so-called weighted $I$-optimality ($\Phi(\M) = \sum_{i=1}^n a_i \f_i' \M^{-1}\f_i$ for given positive weights $a_i$) also belongs to the $\Phi_{\mathfrak{B}}$ class and it can be addressed in a similar manner.} ($K=1$, $\B_1 = (\f_1, \ldots, \f_n)$), $MV$-optimality ($K=m$, $\B_\ell = \e_\ell$ for $\ell \in \{1{:}m\}$), and $G$-optimality ($K=n$, $\B_\ell = \f_\ell$ for $\ell \in \{1{:}n\}$).

$\Phi_\Bf$ criteria are crucial for our MILP formulation, as they involve terms that are linear in the covariance matrix $\Sigmab = \M^{-1}(\db)$ of the least-squares estimator $\widehat{\betab}$. We explicitly reformulate these criteria in terms of the covariance matrix as $\Psi_\Bf(\Sigmab)=\Phi_\Bf(\Sigmab^{-1}) = \max_{\ell=1, \ldots, K} \tr(\B_\ell' \Sigmab \B_\ell)$, where $\Sigmab \in \Sbb^m_{++}$, which can be extended to the entire linear space $\Sbb^m$ by simply setting $\Psi_\Bf(\Sigmab) = \max_{\ell=1, \ldots, K} \tr(\B_\ell' \Sigmab \B_\ell)$ for any $\Sigmab \in \Sbb^m$. This reformulation allows for an optimal design problem formulation that is linear in the covariance matrix $\Sigmab$. However, this is not sufficient to make the problem linear: it merely moves the nonlinearity from the objective function to the constraints of the optimization problem because of the nonlinear relationship $\Sigmab = \M^{-1}(\db)$. Nevertheless, by applying McCormick relaxation, we can address the newly introduced nonlinearity in the constraints, thus ultimately arriving at a linear program, as detailed in Section \ref{sMILP}. This also explains why our approach cannot be extended to $D$-optimality: the problem of $D$-optimal design apparently cannot be recast to involve only the aforementioned specific type of nonlinearity.

In the remainder of this paper, we denote the elements of a covariance matrix $\Sigmab$ by $c_{jk}$ and the elements of a $\Phi_\Bf$-optimal covariance matrix $\Sigmab^*$ by $c^*_{jk}$.

\section{MILP formulation}
\label{sMILP}

The problem of $\Phi_\Bf$-optimal design described in the previous section can be formulated as
\begin{align*}
	\min_{\M,\db} \quad & \max_{\ell=1, \ldots, K} \tr(\B_\ell' \M^{-1} \B_\ell) \\
	s.t. \quad &\M = \sum_{i=1}^n d_i \f_i\f_i', \\
	& \M \in \Sbb^m_{++}, \: \db \in \mathcal{D}_N^{(n)}.
\end{align*}
Note that we assume that the set of feasible solutions is nonempty. 
Clearly, this problem can be rewritten in the following form:
\begin{align}
	\min_{\M, \Sigmab, \db} \quad & \max_{\ell=1, \ldots, K} \tr(\B_\ell' \Sigmab \B_\ell) \label{ePhiopt} \\
	s.t. \quad & \M = \sum_{i=1}^n d_i \f_i\f_i', \: \M \Sigmab = \I_m, \notag \\
	& \M, \Sigmab \in \Sbb^m, \: \db \in \mathcal{D}_N^{(n)}. \notag
\end{align}

Since the objective function in \eqref{ePhiopt} is the maximum of linear functions of $\Sigmab$, the only nonlinear part of problem \eqref{ePhiopt} is $\M\Sigmab = \I_m$. However, this condition can be reformulated using the relaxation method proposed by \cite{McCormick}, analogously to \cite{BhelaEA}. The trick, and the key difficulty, lies in finding a priori bounds $L_{jk}$ and $U_{jk}$ for each element $c_{jk}^*$ of any optimal $\Sigmab^* =\M_*^{-1}$: 
\begin{equation}\label{eBounds}
	L_{jk} \leq c_{jk}^* \leq U_{jk}, \quad j,k \in \{1{:}m\},
\end{equation}
and then constructing new variables
\begin{equation}\label{eNewVariables}
	z_{ijk} = d_i c_{jk}, \quad i \in \{1{:}n\};\, j,k \in \{1{:}m\}.
\end{equation}
The condition $\M\Sigmab = \I_m$ can then be expressed as a set of linear equations for the $z_{ijk}$s by applying the substitution $\M = \sum_i d_i \f_i\f'_i$. This condition is linear in the $nm^2$-dimensional real vector $\z$ composed of all variables $z_{ijk}$, and we can express it as $\A\z = \bb$ for an appropriately chosen $\A$ and $\bb$.\footnote{It does not seem that there is a simple formula for $\A$ and $\bb$ as a function of the elements of $\f_1,\ldots,\f_n$;
see Appendix A for the detailed formulation of $\A\z = \bb$.}

However, in this process, the nonlinear conditions $z_{ijk} = d_i c_{jk}$ are introduced. Note that without altering the set of optimal solutions, we can add the following inequalities:
$$\begin{aligned}
	&d_i (c_{jk} - L_{jk}) \geq 0, \\
	&(d_i - 1) (c_{jk} - U_{jk}) \geq 0, \\
	&d_i (c_{jk} - U_{jk}) \leq 0, \\
	&(d_i - 1) (c_{jk} - L_{jk}) \leq 0,
\end{aligned}$$
which can be expressed using the variables $z_{ijk}$ as
\begin{align}
	&z_{ijk} - d_i L_{jk} \geq 0, \label{eMcCormick1} \\
	&z_{ijk} - d_i U_{jk} - c_{jk} + U_{jk} \geq 0, \label{eMcCormick2}\\
	& z_{ijk} - d_i U_{jk} \leq 0, \label{eMcCormick3}\\
	&z_{ijk} - d_i L_{jk} - c_{jk} + L_{jk} \leq 0\label{eMcCormick4}
\end{align}
for all $i \in \{1{:}n\}$ and $j,k \in \{1{:}m\}$.
These inequalities hold for any optimal $\Sigmab^*$ and $\db^*$ because $0 \leq d_{i}^* \leq 1$ and $L_{jk} \leq c_{jk}^* \leq U_{jk}$; thus, they can be added to \eqref{ePhiopt}. Interestingly, by adding the linear conditions expressed in \eqref{eMcCormick1}-\eqref{eMcCormick4}, one can actually \emph{replace} the nonlinear conditions \eqref{eNewVariables} in \eqref{ePhiopt} (as observed by \cite{BhelaEA} in the network topology context): because of the binary nature of $\db$, conditions \eqref{eMcCormick1}-\eqref{eMcCormick4} imply \eqref{eNewVariables}. In particular, if $d_i = 0$, then \eqref{eMcCormick1} and \eqref{eMcCormick3} imply that $z_{ijk} = 0$, and if $d_i = 1$, then \eqref{eMcCormick2} and \eqref{eMcCormick4} imply that $z_{ijk} = c_{jk}$. In each case, $z_{ijk} = d_i c_{jk}$, and thus, \eqref{eNewVariables} is unnecessary.

Therefore, we obtain an equivalent\footnote{In the sense of having the same set of optimal solutions $\db^*$.} formulation of \eqref{ePhiopt}:
\begin{align}
	\min_{\z, \Sigmab, \db} \quad &\max_{\ell=1, \ldots, K} \tr(\B_\ell' \Sigmab \B_\ell) \label{ePhioptLin} \\ 
	s.t. \quad &\A\z=\bb, \: \eqref{eMcCormick1}-\eqref{eMcCormick4},\notag\\
	& \z \in \R^{nm^2}, \: \Sigmab \in \Sbb^m, \: \db \in \mathcal{D}_N^{(n)}, \notag
\end{align}
where all the constraints are linear in the variables $(\z, \Sigmab, \db)$. This is then equivalent to 
\begin{align}
	\min_{\varphi, \z, \Sigmab, \db} \quad &\varphi \label{ePhiWLin} \\
	s.t. \quad &\varphi \geq \tr(\B_\ell' \Sigmab \B_\ell), \: \ell \in \{1{:}K\}, \notag\\
	&\A\z=\bb, \: \eqref{eMcCormick1}-\eqref{eMcCormick4},\notag\\
	& \varphi \in \R, \: \z \in \R^{nm^2}, \: \Sigmab \in \Sbb^m, \: \db \in \mathcal{D}_N^{(n)}, \notag
\end{align}
where even the objective function is linear. Since all the constraints in \eqref{ePhiWLin} are linear or binary,\footnote{Recall that $\mathcal{D}_N^{(n)}$ is a set of binary vectors $\db$ subject to a linear constraint $\1'_n\db=N$.} problem \eqref{ePhiWLin} is a mixed-integer linear program. Therefore, we have expressed the $\Phi_\Bf$-optimal binary design problem in the form of an MILP problem. As special cases, we can obtain the MILP formulations for $A$-, $I$-, $MV$- and $G$-optimal binary design by choosing the corresponding optimality criteria $\Phi_\Bf$. In Appendix A, we provide a precise formulation of \eqref{ePhiWLin} in vector form, which can therefore be used as input for MILP solvers. Nonetheless, to use the reformulation \eqref{ePhiWLin}, we still need to determine the coefficients $L_{jk}$ and $U_{jk}$, which appear in constraints \eqref{eMcCormick1}-\eqref{eMcCormick4}.

\section{Covariance bounds}
\label{sBounds}

The MILP formulation \eqref{ePhiWLin} requires interval bounds \eqref{eBounds} on the elements of $\Sigmab^*$, which we call covariance bounds. The construction of such bounds is a rich problem, interesting not only for the computation of optimal designs but also in and of itself as a potentially useful characteristic of an experimental design problem. In this section, we describe selected strategies for constructing such interval bounds.

\subsection{General $\Phi_\Bf$-optimality}
\label{ssBoundsGeneral}

Our construction of the covariance bounds relies on knowledge of a design that has a nonsingular information matrix. Therefore, let $\db_0 \in \mathcal{D}_N^{(n)}$ have a nonsingular information matrix $\M_0:=\M(\db_0) \in \Sbb^m_{++}$, and let $\Sigmab_0 = \M_0^{-1}$. The more $\Phi_\Bf$-efficient $\db_0$ is, the stronger are the bounds we obtain; thus, in applications, we recommend computing $\db_0$ via a $\Phi_\Bf$-optimization heuristic (e.g., some exchange algorithm; see Chapter 12 in \cite{AtkinsonEA07}).

Let $\alpha := \Phi_\Bf(\M_0)$. Clearly, $\Psi_\Bf(\Sigmab^*) \leq \Psi_\Bf(\Sigmab_0) = \Phi_\Bf(\M_0)$, that is,
\begin{equation}\label{eq:alpha}
	\tr(\B_\ell'\Sigmab^* \B_\ell) \leq \alpha, \: \ell \in \{1{:}K\}.
\end{equation}
We will show that the constraints \eqref{eq:alpha} are sufficient to provide finite bounds on all the elements of $\Sigmab^*$.

\subsubsection{Computational construction}
\label{ssBoundsComp}

Recall the notation $c^*_{jk}:=(\Sigmab^*)_{jk}$ for $j,k \in \{1{:}m\}$. A direct approach to finding the covariance bounds is a computational one: apply mathematical programming to find the smallest/largest value of $(\Sigmab)_{jk}$ over all matrices $\Sigmab \in \mathbb{S}^m_+$ that satisfy \eqref{eq:alpha}. Then, $c^*_{jk}$ must be bounded by these values. Formally, for any $j,k \in \{1{:}m\}$,
\begin{eqnarray}\label{eq:SDPupper}
	c^*_{jk} \quad \leq \quad \max_{\Sigmab} && \e_j' \Sigmab \e_k, \\ 
	s.t. && \tr(\B'_\ell \Sigmab \B_\ell) \leq \alpha, \: \ell \in \{1{:}K\},\nonumber \\
	&& \Sigmab \in \mathbb{S}^{m}_{+};\nonumber
\end{eqnarray}
\begin{eqnarray}\label{eq:SDPlower}
	c^*_{jk} \quad \geq \quad \min_{\Sigmab} && \e_j' \Sigmab \e_k, \\ 
	s.t. && \tr(\B'_\ell \Sigmab \B_\ell) \leq \alpha, \: \ell \in \{1{:}K\}, \nonumber\\
	&& \Sigmab \in \mathbb{S}^{m}_{+}.\nonumber
\end{eqnarray}

Optimization problems \eqref{eq:SDPupper} and \eqref{eq:SDPlower} are (continuous, not discrete) semidefinite programming (SDP) problems; that is, they are easy to solve using readily available and efficient SDP solvers. Moreover, the above bounds are, by definition, the strongest ones that can be constructed using only the inequalities \eqref{eq:alpha}. 

In the case of the lower diagonal bounds, problem \eqref{eq:SDPlower} can be analytically solved to arrive at the simple bound $c_{jj}^* \geq 0$. This is because $\e_j^T \Sigmab \e_j \geq 0$ for any $\Sigmab \in \Sbb^m_+$, and the objective function value $\e_j^T \Sigmab \e_j = 0$ is attained for the feasible solution $\Sigmab = \0_{m \times m}$. On the other hand, the upper bounds on the diagonal elements of $\Sigmab^*$ as well as the lower and upper bounds on the nondiagonal elements of $\Sigmab^*$ generally depend on the optimality criterion and the value of $\alpha$.
\bigskip

To find bounds for the entire $\Sigmab^*$, a distinct pair of problems \eqref{eq:SDPupper} and \eqref{eq:SDPlower} must be solved for each $c_{jk}^*$. This therefore requires solving $m(m+1)$ semidefinite programs\footnote{More precisely, $m(m+1) - m = m^2$ programs, because the $m$ lower diagonal bounds have analytical solutions.}, which in some cases may be inconvenient. In addition, the finiteness of the optimal values of \eqref{eq:SDPupper} and \eqref{eq:SDPlower} is not immediately apparent; however, these values must be finite to be of any utility. We therefore also provide analytical bounds on the $c^*_{jk}$s that guarantee the finiteness of \eqref{eq:SDPupper} and \eqref{eq:SDPlower} (see Theorem \ref{tSDP}) and do not require the numerical solution of auxiliary optimization problems.

\subsubsection{Analytical construction}
\label{ssBoundsAnal}

First, we provide two simple constructions of the bounds on the nondiagonal covariances $c^*_{jk}$ based on the bounds on the variances $c_{jj}^*$. Because $\Sigmab^*$ is positive semidefinite, Sylvester's criterion for positive semidefinite matrices implies that $c_{jj}^* \geq 0$ and $c_{jj}^*c_{kk}^* - (c_{jk}^*)^2 \geq 0$, i.e.,
\begin{equation}\label{eBoundGeneral2}
	|c_{jk}^*| \leq \sqrt{c_{jj}^* c_{kk}^*}, \quad j,k \in \{1{:}m\}.
\end{equation}
Furthermore, the inequality of the geometric and arithmetic means provides weaker but linear bounds, which can also be useful:
\begin{equation}\label{eBoundGeneral3}
	|c_{jk}^*| \leq \frac{1}{2}(c_{jj}^*+c_{kk}^*), \quad j,k \in \{1{:}m\}. 
\end{equation}
Therefore, to find both lower and upper bounds on the $c^*_{jk}$s for all $j,k \in \{1{:}m\}$, we need only to construct finite upper bounds on the variances $c^*_{jj}$ for all $j \in \{1{:}m\}$ or on the sums $c_{jj}^*+c_{kk}^*$ for all $j,k \in \{1{:}m\}$.
\bigskip

The key mathematical result for the analytical bounds is the following lemma. The proof of Lemma \ref{lemma1} and all other nontrivial proofs are deferred to Appendix B.

\begin{lemma}\label{lemma1}
	Let $\Sigmab$ be any $m \times m$ nonnegative definite matrix such that $\tr(\B_\ell'\Sigmab \B_\ell) \leq \alpha$ for all $\ell \in \{1{:}K\}$. Let $\w=(w_1,\ldots,w_K)' \in \mathbb{R}^K$ be a vector with nonnegative components summing to $1$, and let $\N(\w)=\sum_{\ell=1}^K w_\ell \B_\ell \B'_\ell$. Assume that $\X$ is an $m \times r$ matrix such that $\C(\X) \subseteq \C(\N(\w))$. Then,
	\begin{equation}\label{eq:Lemma1}
		\tr(\X' \Sigmab \X) \leq \alpha \lambda_{\max}(\X'\N^+(\w)\X).
	\end{equation}
\end{lemma}

Because  $\tr(\B_\ell'\Sigmab^* \B_\ell) \leq \alpha$ for all $\ell \in \{1{:}K\}$, we can use Lemma \ref{lemma1} with various choices for $\X$ and $\w$ to obtain bounds on the elements of $\Sigmab^*$. Let $j,k \in \{1{:}m\}$, $j \neq k$. Among the large variety of possibilities, we will use only the matrices $\X=\e_j$, which directly provide bounds on $c_{jj}^*$, and $\X=\E_{jk}:=(\e_j,\e_k)$, which provide bounds on $c_{jk}^*$ due to inequality \eqref{eBoundGeneral3}. For such $\X$, any $\w \in \R^k$ that satisfies $\sum_\ell w_\ell = 1$, $\w \geq \0$, and $\C(\X) \subseteq \C(\N(\w))$ can be used. 

It is then of interest to examine choices for $\w$. Interestingly, such a vector $\w$ can be viewed as an approximate design for the artificial model $\y_\ell = \B_\ell'{\boldsymbol \theta} + {\boldsymbol \varepsilon}_\ell$, with $s_\ell$-dimensional observations $\y_\ell$, design space $\{1{:}K\}$ and elementary information matrices $\B_1 \B_1', \ldots, \B_K\B_K'$. The corresponding information matrix of $\w$ is then $\sum_\ell w_\ell \B_\ell \B_\ell'$, which is exactly $\N(\w)$. Therefore, we refer to such vectors $\w$ as (approximate) designs. In particular, we consider the following choices:
\begin{itemize}
	\item The \textbf{uniform design} $\w^{(u)}=\1_K/K$; note that $\N(\w^{(u)})=\frac{1}{K}\B\B'$.
	\item The design $\w^{(j+)}$ with components $w_\ell^{(j+)}=\|\h^{(j+)}_{\ell}\|/S_j$ for all $\ell \in \{1{:}K\}$, where $\h^{(j+)}_{1},\ldots,\h^{(j+)}_{K}$ are vectors of dimensions $s_1,\dots,s_K$ such that
	\begin{equation*}
		((\h^{(j+)}_{1})',\ldots,(\h^{(j+)}_{K})')'=\B^+\e_j
	\end{equation*} and $S_j = \sum_{\ell} \|\h^{(j+)}_{\ell}\|$. We call $\w^{(j+)}$ the \textbf{Moore--Penrose approximate design for $\e_j$} for the artificial model.
	\item The design $\w^{(j*)}$ that minimizes $\e_j'\N^+(\w)\e_j$ in the class of all approximate designs $\w$ (i.e., $\w \geq \0_K$, $\1_K'\w=1$) satisfying $\e_j \in \C(\N(\w))$. We call $\w^{(j*)}$ the \textbf{$\e_j$-optimal approximate design} for the artificial model. 
	\item $\w^{(jk+)}:=\frac{1}{2}\left(\w^{(j+)}+\w^{(k+)}\right)$ and $\w^{(jk*)}:=\frac{1}{2}\left(\w^{(j*)}+\w^{(k*)}\right)$.
\end{itemize}

The uniform design is the most straightforward choice. A more refined choice is the Moore--Penrose approximate design, which accounts for the desirability of having a ``small'' matrix $\N^+(\w)$. The $\e_j$-optimal approximate design actually optimizes the right-hand side of \eqref{eq:Lemma1} for $\X=\e_j$ to make it as small as possible; such designs are theoretically interesting but not necessarily practically desirable. Finally, the designs $\w^{(jk+)}$ and $\w^{(jk*)}$ are constructed such that $\X = \E_{jk}$ satisfies $\C(\X) \subseteq \C(\N(\w))$. Note that although we use approximate designs as an auxiliary tool for computing covariance bounds, \textit{optimal} approximate designs are needed only for the computation of $\w^{(j*)}$ and $\w^{(jk*)}$.

\begin{lemma}\label{lem:C}
	Let $j \in \{1{:}m\}$. The column space of each of the matrices $\N(\w^{(u)})$, $\N(\w^{(j+)})$, and $\N(\w^{(j*)})$ contains $\e_j$. Let $j,k \in \{1{:}m\}$, $j \neq k$. The column space of each of the matrices $\N(\w^{(u)})$, $\N(\w^{(jk+)})$, and $\N(\w^{(jk*)})$ contains the column space of the matrix $\E_{jk}$.
\end{lemma}

According to Lemma \ref{lem:C},  for $t \in \{j,k\}$, we can use Lemma \ref{lemma1} with $\w=\w^{(u)}$, $\w=\w^{(t+)}$, and $\w=\w^{(t*)}$ and with $\X=\e_t$ and apply bound \eqref{eBoundGeneral2} to obtain the following theorem:

\begin{theorem}[Type I covariance bounds]\label{tTypeI}
	For any $j,k \in \{1{:}m\}$, we have
	\begin{eqnarray}
		|c_{jk}^*| &\leq& \alpha K [(\B\B')^{-1}_{jj}(\B\B')^{-1}_{kk}]^{1/2}, \label{eq:Bnd13} \\
		|c_{jk}^*| &\leq& \alpha [\N^+_{jj}(\w^{(j+)})\N^+_{kk}(\w^{(k+)})]^{1/2}, \label{eq:Bnd12} \\
		|c_{jk}^*| &\leq& \alpha [\N^+_{jj}(\w^{(j*)}) \N^+_{kk}(\w^{(k*)})]^{1/2}, \label{eq:Bnd11}
	\end{eqnarray}
	where $(\B\B')^{-1}_{ij}$ and $\N^+_{ij}(\w)$ represent the elements of matrices $(\B\B')^{-1}$ and $\N^+(\w)$ with coordinates $(i, j)$, respectively.
\end{theorem}

Similarly, for $j,k$, $j \neq k$, we can use Lemma \ref{lemma1} with $\w=\w^{(u)}$, $\w=\w^{(jk+)}$, $\w=\w^{(jk*)}$, and $\X=\E_{jk}$ and apply bound \eqref{eBoundGeneral3}. We thus obtain the following theorem.

\begin{theorem}[Type II covariance bounds]\label{tTypeII}
	For any $j,k \in \{1{:}m\}$, $j \neq k$, we have
	\begin{eqnarray}
		|c_{jk}^*| &\leq& \frac{\alpha}{2} K \lambda_{\max}(\E'_{jk}(\B\B')^{-1}\E_{jk}), \label{eq:Bnd23}\\
		|c_{jk}^*| &\leq& \frac{\alpha}{2} \lambda_{\max}(\E'_{jk}\N^+(\w^{(jk+)})\E_{jk}),  \label{eq:Bnd22}\\
		|c_{jk}^*| &\leq& \frac{\alpha}{2} \lambda_{\max}(\E'_{jk}\N^+(\w^{(jk*)})\E_{jk}). \label{eq:Bnd21}
	\end{eqnarray}
\end{theorem}

The bounds provided by the above theorems can be computed analytically or by means of standard numerical linear algebra and relatively simple optimization. They also guarantee the finiteness of the SDP-based bounds (e.g., using \eqref{eq:Bnd13} and \eqref{eq:Bnd23}):

\begin{theorem}[Finiteness of the SDP-based bounds]\label{tSDP}
	Let $\db_0 \in \mathcal{D}_N^{(n)}$ have a nonsingular information matrix $\M_0:=\M(\db_0) \in \Sbb^m_{++}$, and let $\alpha = \Phi_\Bf(\M_0)$. Then, the optimal values of the optimization problems \eqref{eq:SDPupper} and \eqref{eq:SDPlower} are finite.
\end{theorem}

We emphasize that the analytical bounds from Theorems \ref{tTypeI} and \ref{tTypeII} are never stronger than the computational bounds \eqref{eq:SDPupper} and \eqref{eq:SDPlower}. However, they are often very simple, and for some criteria, they even coincide with the SDP-based bounds, as demonstrated below.

\subsection{$A$-, $I$-, $MV$-, and $G$-optimality}
\label{ssBoundsCrit}

As shown in Section \ref{ssBoundsComp}, we have $c_{jj}^* \geq 0$ for $j \in \{1{:}m\}$. We therefore further examine only the remaining bounds. Table \ref{tblCriteria2} provides an overview of the results for $A$-, $I$-, $MV$-, and $G$-optimality, and the following subsections give detailed explanations of these bounds.

\begin{table}[h]
	\centering
	\begin{tabular}{c | c c |c c}
		\hline
		\multirow{2}{*}{$\Phi$} & \multicolumn{2}{c|}{diagonal: $c_{jj}^*\leq$} & \multicolumn{2}{c}{nondiagonal: $\lvert c_{jk}^*\rvert\leq$} \\
		& computational & analytical & computational & analytical  \\ \hline
		$A$ & $\alpha$ & $\alpha$ & $\alpha/2$ & $\alpha/2$ \\
        \multirow{2}{*}{$I$} & \multirow{2}{*}{$\star$} &  \multirow{2}{*}{$\alpha (\F\F')^{-1}_{jj}$} & \multirow{2}{*}{$\star$} & $ \alpha\min\left\{[(\F\F')^{-1}_{jj}(\F\F')^{-1}_{kk}]^{1/2},\right.$  \\
        & & & &  $\left. \lambda_{\max}(\E'_{jk}(\F\F')^{-1}\E_{jk})/2\right\} $  \\
        $MV$ & $\alpha$ & $\alpha$ & $\alpha$ & $\alpha$ \\
        $G$ & $\star$ & Section \ref{subsubsec:G} & $\star$ & Section \ref{subsubsec:G}
        \\
		\hline
	\end{tabular}
	\caption{Overview of the covariance bounds for the cases of $A$-, $I$-, $MV$-, and $G$-optimality, as obtained from solutions to \eqref{eq:SDPupper} and \eqref{eq:SDPlower} (computational) and from Theorems \ref{tTypeI} and \ref{tTypeII} (analytical). A star $\star$ denotes that we are not aware of an explicit solution to \eqref{eq:SDPupper} and \eqref{eq:SDPlower}. In the case of $G$-optimality, the reader is referred to Section \ref{subsubsec:G} for the analytical bounds, as they do not lend themselves to a succinct description.}
	\label{tblCriteria2}
\end{table}

\subsubsection{$A$-optimality}

For $A$-optimality ($K=1$, $\B_1 = \B = \I$), $\alpha = \tr(\M^{-1}_0)$, and $j,k \in \{1{:}m\}$, bounds \eqref{eq:Bnd13}-\eqref{eq:Bnd11} and \eqref{eq:Bnd23}-\eqref{eq:Bnd21} reduce to 
\begin{eqnarray}
	c^*_{jj} &\leq& \alpha, \label{eq:BndAjj}\\
	|c^*_{jk}| &\leq& \frac{\alpha}{2}, \: j \neq k.\label{eq:BndAjk}
\end{eqnarray}
By solving the SDP formulations \eqref{eq:SDPupper} and \eqref{eq:SDPlower}, we again obtain only \eqref{eq:BndAjj} and \eqref{eq:BndAjk}; for the diagonal elements, we have already proven that $c_{jj}^* \leq \alpha$, so it is sufficient to find a $\Sigmab \in \mathbb{S}^m_+$ with $\tr(\Sigmab) \leq \alpha$ and $c_{jj} = \alpha$. These conditions are satisfied for $\Sigmab = \alpha\e_j\e_j'$. For the nondiagonal elements, the maximum is attained for $\Sigmab = \alpha (\e_j + \e_k)(\e_j + \e_k)'/2$, and the minimum is attained for $\Sigmab = \alpha (\e_j - \e_k)(\e_j - \e_k)'/2$.

\subsubsection{$I$-optimality}

For $I$-optimality ($K=1$, $\B_1 = \B = \F := (\f_1,\ldots,\f_n)$), $\alpha = \tr(\F'\M^{-1}_0\F)$, and $j,k \in \{1{:}m\}$, we obtain the following bounds from \eqref{eq:Bnd13}-\eqref{eq:Bnd11} and \eqref{eq:Bnd23}-\eqref{eq:Bnd21}:
\begin{eqnarray}
	|c^*_{jk}| &\leq& \alpha [(\F\F')^{-1}_{jj}(\F\F')^{-1}_{kk}]^{1/2}, \label{eq:BndIjj}\\
	|c^*_{jk}| &\leq& \frac{\alpha}{2} \lambda_{\max}(\E'_{jk}(\F\F')^{-1}\E_{jk}), \: j \neq k.\label{eq:BndIjk}
\end{eqnarray}

Neither of these bounds dominates the other. For instance, let $(\F\F')^{-1}=\I_2$, $j=1$ and $k=2$. Then, \eqref{eq:BndIjj} becomes $|c^*_{12}| \leq \alpha$, and \eqref{eq:BndIjk} becomes $|c^*_{12}| \leq \alpha/2$. In contrast, if $(\F\F')^{-1}$ is a $2 \times 2$ diagonal matrix with 100 and 1 on the diagonal, then  \eqref{eq:BndIjj} is $|c^*_{12}| \leq 10\alpha$ and  \eqref{eq:BndIjk} is $|c^*_{12}| \leq 50\alpha$.

\subsubsection{$MV$-optimality}\label{subsubsec:MV}

For $MV$-optimality ($K=m$, $\B_\ell = \e_\ell$, $\ell \in \{1{:}m\}$, $\B=\I$), $\alpha = \max_{\ell} \e'_\ell \M^{-1}_0 \e_\ell$, let $j,k \in \{1{:}m\}$ and note that $\w^{(j+)}=\w^{(j*)}=\e_j$. Therefore, bounds \eqref{eq:Bnd12}, \eqref{eq:Bnd11}, \eqref{eq:Bnd22} and \eqref{eq:Bnd21} all lead to the same conclusion:
\begin{equation}\label{eq:BndMV}
	|c^*_{jk}| \leq \alpha.
\end{equation}
Bounds \eqref{eq:Bnd13} and \eqref{eq:Bnd23} lead to weaker constraints of $|c^*_{jk}| \leq m \alpha$ and $|c^*_{jk}| \leq m \alpha/2$, respectively.

Again, the SDP formulations \eqref{eq:SDPupper} and \eqref{eq:SDPlower} cannot improve upon \eqref{eq:BndMV}. As in the case of $A$-optimality, this can be easily proven: both the diagonal and nondiagonal maxima for $MV$-optimality are attained at $\Sigmab = \alpha\1_m \1_m'$, and the nondiagonal minimum is attained at $\Sigmab = \alpha (\e_j - \e_k)(\e_j - \e_k)'$.

\subsubsection{$G$-optimality}\label{subsubsec:G}

For $G$-optimality ($K=n$, $\B_\ell = \f_\ell$, $\ell \in \{1{:}n\}$, $\B=\F= (\f_1,\ldots,\f_n)$), the situation becomes more complex and more interesting. Let $\alpha = \max_{\ell} \f'_\ell \M^{-1}_0\f_\ell$.\footnote{It may be efficient to compute $\M_0=\M(\db_0)$ via a heuristic for $D$-optimality, as these algorithms are typically faster and more readily available than those for $G$-optimality. We note that $D$- and $G$-optimal designs tend to be close because they coincide in approximate design theory.} Note that for $G$-optimality, the primary (original) and the artificial models coincide; thus, the auxiliary approximate design $\w$ on $\{1{:}K\}$ is in fact the approximate design on the original design space $\{1{:}n\}$, and $\N(\w)$ is exactly the standard information matrix of an approximate design $\N(\w)=\sum_{i=1}^n w_i \f_i\f'_i$. Then,
\begin{itemize}
	\item bounds \eqref{eq:Bnd13} and \eqref{eq:Bnd23} become $|c^*_{jk}| \leq n\alpha [(\F\F')^{-1}_{jj}(\F\F')^{-1}_{kk}]^{1/2}$ for any $j, k$ and $|c_{jk}^*| \leq n \alpha  \lambda_{\max}(\E'_{jk}(\F\F')^{-1}\E_{jk}) / 2$
	for $j \neq k$;
	\item bounds \eqref{eq:Bnd12} and \eqref{eq:Bnd22} require $\w^{(t+)}$ for each $t \in \{1{:}n\}$, which is straightforward to calculate using $\h^{(t+)} = \F^+ \e_t$ and
	$$\w^{(t+)}= (|h_1^{(t+)}|, \ldots, |h_n^{(t+)}|)' / \sum_{i=1}^n |h_i^{(t+)}|;$$
	\item bounds \eqref{eq:Bnd11} and \eqref{eq:Bnd21} require computing $\w^{(t*)}$, which in this case is the standard $\e_t$-optimal approximate design; it is known that this design can be efficiently computed via continuous linear programming (see \cite{HarmanJurik}).
\end{itemize}

\subsection{Further improvements}\label{ssImpBnds}

Let $j,k \in \{1{:}m\}$. A general strategy for the construction of interval bounds $[L_{jk}, U_{jk}]$ for the optimal $c_{jk}^*$ is as follows. First, find appropriate $m \times m$ matrices $\Hb_t$ and numbers $\gamma_t$ for $t \in T$, where $T$ is an index set, such that we can be sure that $\Sigmab^*$ satisfies the constraints
\begin{equation}\label{eq:genlincon}
	\tr(\Hb_t \Sigmab^*) \leq \gamma_t, \: t \in T.
\end{equation}
Second, construct $L_{jk}$ and $U_{jk}$ such that $[L_{jk}, U_{jk}]$ contains the values $(\Sigmab)_{jk}=\e_j'\Sigmab \e_k$ for any $\Sigmab \in \Sbb^m_{++}$ satisfying \eqref{eq:genlincon}. In Section \ref{ssBoundsGeneral}, we applied this strategy with $T=\{1{:}K\}$, $\Hb_t=\B_t\B_t'$, and $\gamma_t\equiv\alpha$. However, constraints of the type \eqref{eq:genlincon} can be added to those from Section \ref{ssBoundsGeneral} to obtain better theoretical covariance bounds. They can also be added to the SDP formulations \eqref{eq:SDPupper} and \eqref{eq:SDPlower} to algorithmically provide better bounds.

For instance, a stronger lower bound for $c_{jj}^*$, $j \in \{1{:}m\}$, can be found by computing the $\e_j$-optimal approximate design for the primary model, i.e., by minimizing $\e_j'\M^+(\w)\e_j$ under the constraint $\e_j \in \C(\M(\w))$, where $\w \in \R^n$ ($w_i \geq 0$ for all $i$ and $\sum_i w_i = 1$) is an approximate design for the original model. Let $t_j^*$ be the minimal value of $\e_j'\M^+(\w)\e_j$. Clearly, for any $\db$ with a nonsingular information matrix, we then have $\e_j'\M^{-1}(\db)\e_j = N^{-1} \e_j'\M^{-1}(\db/N)\e_j \geq t_j^*/N$. It follows that $c_{jj}^* \geq t_j^*/N$, which is always stronger than our standard bound $c_{jj}^* \geq 0$. However, unlike $c^*_{jj} \geq 0$, obtaining these $m$ bounds requires solving $m$ $\e_j$-optimal approximate design problems by means of, e.g., linear programming (\cite{HarmanJurik}). 
A similar approach can be applied to the nondiagonal elements. Nevertheless, we have decided not to explore these possibilities in this paper and instead leave them for further research.

\section{Extensions}\label{sec:Extend}

In this section, we provide guidelines for more complex optimal design problems that can be solved in a straightforward way via the proposed MILP approach. 

\subsection{Designs with replications}
\label{ssRepeated}

If we have a $\Phi_\Bf$-optimal design problem in which we allow for up to $N_i$ replicated observations at the design point $i \in \{1{:}n\}$, an MILP formulation can be obtained by replicating the regressor $\f_i$ exactly $N_i$ times and then using \eqref{ePhiWLin}. Setting $N_i=N$ for each $i \in \{1{:}n\}$ allows us to compute optimal designs for the classical problem, i.e., with a single constraint representing the total number $N$ of replications at all design points combined, which is demonstrated in Section \ref{ssQuadratic}. 

\subsection{Multiple design constraints}
\label{ssDesignConst}

In some applications, we need to restrict the set of permissible designs to a set smaller than $\mathcal{D}^{(n)}_N$. Such additional constraints on $\db$ can correspond to safety, logistical or resource restrictions. Many such constraints are linear (see, e.g., \cite{Harman14}, \cite{SagnolHarman15} and \cite{HarmanEA16} for examples), and they can be directly included in \eqref{ePhiWLin}. Another class of linear constraints corresponds to the problem of optimal design augmentation, which entails finding an optimal design when specified trials must be performed or have already been performed (see Chapter 19 in \cite{AtkinsonEA07}). Constraints of this kind can also be incorporated into the MILP formulation in a straightforward way ($d_i \geq N_i$, where $N_i \in \{0,1,\ldots\}$ is the number of trials already performed at point $i$), but a design $\db_0$ satisfying the required constraints must be used to construct the covariance bounds. However, the practical nature of the constraints usually allows for simple selection of such an initial $\db_0$.
MILP with additional design constraints is used in Section \ref{ssQuadratic} to obtain a ``space-filling'' optimal design and in Section \ref{ssNonlinear} to find a cost-constrained optimal design (see Sections \ref{ssQuadratic} and \ref{ssNonlinear} for the formal definitions of such designs).

\subsection{Optimal designs under covariance constraints}
\label{ssConstraints}

A unique feature of the proposed approach is that we can add to \eqref{ePhiWLin} any linear constraints on the elements of the variance matrix $\Sigmab$, and the resulting problem will still be a mixed-integer linear program, whereas the available heuristics for computing optimal designs do not allow for such variance--covariance constraints. For instance, we can specify that the absolute values of all covariances should be less than a selected threshold ($|c_{jk}| \leq c$ for $j \neq k$), with the aim of finding a design that is optimal among those that are ``close'' to orthogonal (or even exactly orthogonal). Similarly, we can add the condition that the sum of all variances of the least-squares estimators should be less than some number ($\sum_j c_{jj} \leq c$), with the aim of, e.g., finding the $MV$-optimal design among all designs that attain at most a given value of $A$-optimality. More generally, we can compute a $\Phi_\Bf$-optimal design that attains at most a selected value of a different criterion from the same class. 

As in Section \ref{ssDesignConst}, a design $\db_0$ that satisfies the required constraints must be used to construct the covariance bounds---which may not be easy to find for constraints such as $|c_{jk}| \leq c$. Fortunately, one class of constraints has such designs $\db_0$ readily available: those that require the design to attain a certain value of a different optimality criterion from the $\Phi_\Bf$ class. For instance, we can find the $MV$-optimal design among all the designs that satisfy $\sum_j c_{jj} \leq c$ (i.e., $\Phi_A(\M(\db)) \leq c$: a limit on the value of $A$-optimality). Here, $\db_0$ can be chosen as either the $A$-optimal design or a design obtained via a heuristic for $A$-optimality. Such $\db_0$ then allows for any $c$ such that $c \geq \Phi_A(\M(\db_0))$, which captures all reasonable choices for $c$.\footnote{A lower $c$ would mean that we seek a design that is better with respect to $A$-optimality than either the $A$-optimal design itself or some best known design.}
The applicability of this approach is also demonstrated in Section \ref{ssQuadratic}.

\subsection{Infinite design spaces}

We focus on finite design spaces because integer programming approaches are unsuitable for continuous design spaces (as there would be an infinite number of variables in the latter case). Finiteness of the design set is common in both experimental design and related disciplines (e.g., \cite{Mitchell}, \cite{AtkinsonDonev}, \cite{GhoshEA08}, \cite{BaileyCameron}, \cite{Todd16}). Moreover, even if the theoretical design space is infinite, it can be discretized, which is a typical technique in design algorithms (e.g., \cite{Fedorov89}, \cite{MeyerNachtsheim}). This often results in only a negligible loss in the efficiency of the achievable design. Although some heuristics can handle continuous design spaces directly, they are still only heuristics and, as such, risk providing a design that is worse than one obtained by applying a discrete optimizer to a discretized design space (cf. \cite{HarmanEA21}). 

Nevertheless, for the case of continuous design spaces, the MILP solution can be very useful for finding the best design on a small support predetermined by, for instance, some approximate optimal design algorithm, such as that of \cite{HarmanEA21}.
This is demonstrated in Section \ref{ssNonlinear}.

\subsection{Nonlinear models}

Because our results rely on the additive form of the information matrix, they can be straightforwardly extended to the usual generalizations by simply adapting the formula for the information matrix. For a nonlinear regression $y(\x) = \h(\x, \betab) + \varepsilon_\x$ with Gaussian errors, the usual local optimality method can be used (see, e.g., \cite{PronzatoPazman}, Chapter 5), resulting in the information matrix
$$
\M(\db) = \sum_{i=1}^n d_i \left.\frac{\partial \h(\x_i, \betab)}{\partial \betab} \left(\frac{\partial \h(\x_i, \betab)}{\partial \betab} \right )'\right\vert_{\betab = \betab_0},
$$
where $\betab_0$ is a nominal parameter that is assumed to be close to the true value of $\betab$. This approach is demonstrated in Section \ref{ssNonlinear}. Generalized linear models can be dealt with in a similar manner; see \cite{KhuriEA} and \cite{AtkinsonWoods}. In other words, if we use the so-called approach of locally optimal design for nonlinear models, the computational methods are equivalent to those of standard optimal design for linear models.

\section{Numerical study}
\label{sStudy}

All computations reported in this section were performed on a personal computer with the 64-bit Windows 10 operating system and an Intel i5-6400 processor with 8 GB of RAM. Our algorithm was implemented in the statistical software \texttt{R}, the mixed-integer linear programs were solved using the \texttt{R} implementation of the Gurobi software (\cite{Gurobi}), and the MISOCP computations of the optimal designs were performed using the \texttt{OptimalDesign} package (\cite{OptimalDesign}), ultimately also calling the Gurobi solver. The SDP-based bounds \eqref{eq:SDPupper} and \eqref{eq:SDPlower} were solved for via convex programming using the \texttt{CVXR} package (\cite{CVXR}). The \texttt{R} codes for applying our MILP approach are available at \url{http://www.iam.fmph.uniba.sk/ospm/Harman/design/}, and the authors also plan to implement the MILP algorithm in the \texttt{OptimalDesign} package to make it available in a more user-friendly manner. 

Naturally, all the reported running times of the MILP approach include the time it took to compute the relevant bounds via the SDP formulations \eqref{eq:SDPupper} and \eqref{eq:SDPlower} or via the analytical formulas in Section \ref{ssBoundsAnal}. We also note that in all the following applications of our MILP formulation, we used the analytical bounds \eqref{eq:BndAjj}, \eqref{eq:BndAjk} and \eqref{eq:BndMV} for $A$- and $MV$-optimality because they are equivalent to the SDP bounds \eqref{eq:SDPupper} and \eqref{eq:SDPlower} but less time consuming to compute. For $I$- and $G$-optimality, we considered both approaches, and we selected the one that tended to result in faster computation of the entire MILP algorithm: the SDP bounds for $I$-optimality and the analytical bounds for $G$-optimality. Note that although the SDP bounds for $G$-optimality tend to result in shorter total computation times for larger models, the analytical version enabled overall faster MILP computations for the models considered in this paper. The initial designs $\db_0$ in the case of $G$-optimality were computed using the KL exchange algorithm (see Chapter 12 of \cite{AtkinsonEA07}) for $D$-optimality; for all other criteria, the KL exchange algorithm for $A$-optimality was used.

\subsection{Quadratic regression}
\label{ssQuadratic}

Let us demonstrate the MILP algorithm on the quadratic regression model
\begin{equation}\label{eModelQuad}
	y_i = \beta_1 + \beta_2 x_i + \beta_3 x_i^2 + \varepsilon_i,
\end{equation}
where the $x_i$ values lie in the domain $[-1,1]$ equidistantly discretized to $n$ points, i.e., $x_i \in \{-1, -1 + 2/(n-1), \ldots, 1 - 2/(n-1), 1\}$. The exact optimal designs for model \eqref{eModelQuad} have been extensively studied in the past (e.g., \cite{GaffkeKrafft}, \cite{ChangEA}, \cite{Imhof00}, \cite{Imhof}), but for some criteria, sample sizes and design spaces, there are still no known analytical solutions.

\begin{figure}[h]%
	\centering
	\includegraphics[width=0.95\textwidth]{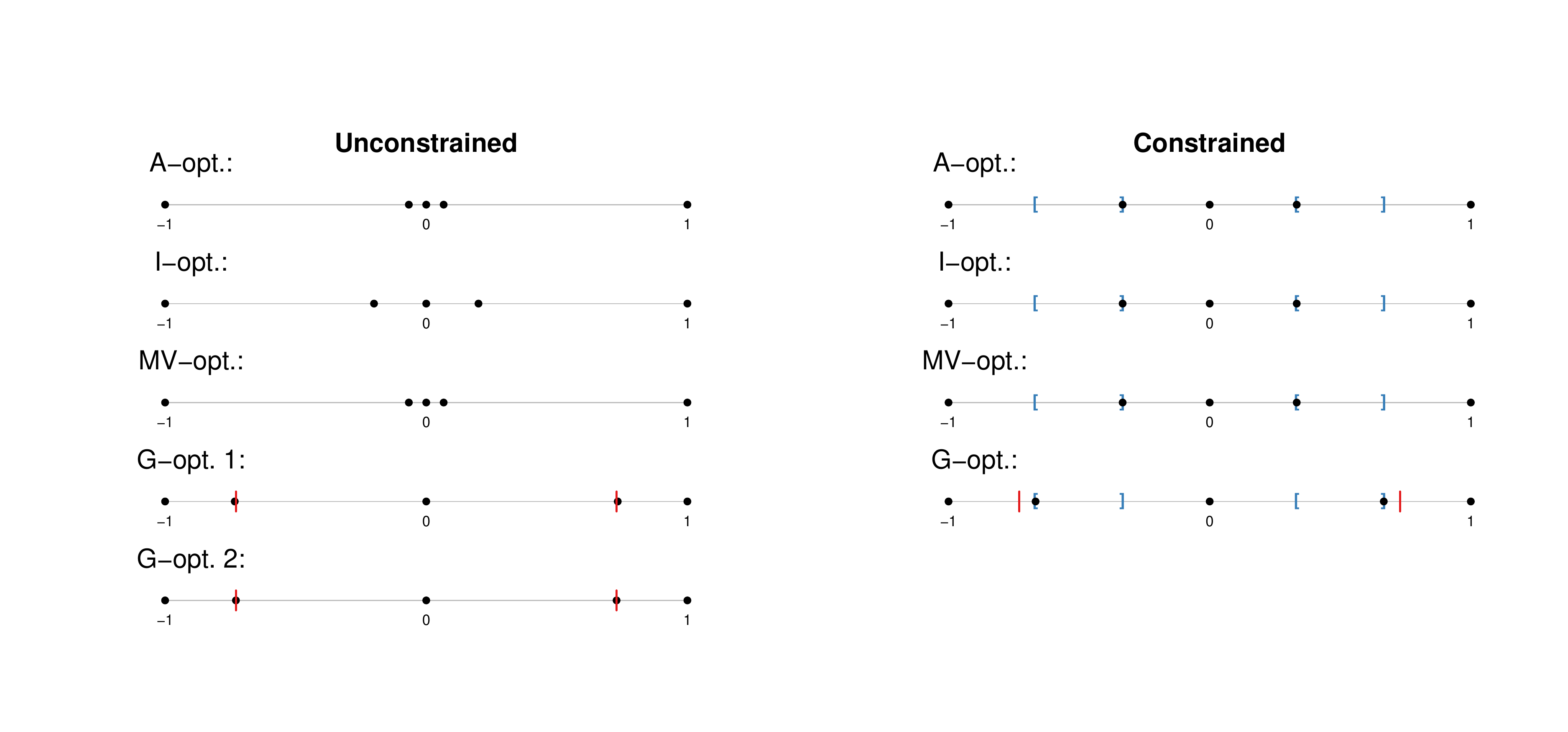}
	\vspace{-20pt}
	\caption{Optimal designs for the quadratic regression model \eqref{eModelQuad}. The dots represent the values of $x_i \in [-1,1]$ for which $d_i = 1$. The designs in the left column are unconstrained, whereas the designs in the right column are constrained by \eqref{eConst1}, with the constraint limits indicated by blue braces [,]. The design ``G-opt. 1'' is for the original (discretized) model, and ``G-opt. 2'' is the design for the model with added points $x=\pm\gamma$; the optimal support points $\gamma$ reported by \cite{Imhof} are indicated by red lines.}\label{fQRall}
\end{figure}

To illustrate the MILP computational approach, we select $n=31$ design points and $N=5$ observations, and we compute the $A$-, $I$-, $MV$- and $G$-optimal designs. The running times of these computations are given in the first row of Table \ref{tblQRall}, and the resulting designs are illustrated in Figure \ref{fQRall} (left). The running times show that the optimal design calculations using MILP strongly depend on the selected optimality criterion.

\begin{table}[h]
	\centering
	\begin{tabular}{c c c c c}
		\hline
		Criterion & A & I & MV & G \\
		Binary & 3.99 (0.26) & 11.75 (1.18) & 2.59 (0.33) & 36.62 (0.90) \\
		Replicated & 74.98 (2.67) & 214.57 (13.66) & 64.44 (4.00) & 487.56 (61.59) \\
		\hline
	\end{tabular}
	\caption{Running times of the MILP computations of optimal designs for the quadratic regression model \eqref{eModelQuad} ($N=5$, $n=31$, $m=3$) for binary designs (first row) and for designs with allowed replications (second row). Each algorithm was run 10 times, and the means and standard deviations of these runs are reported in the form ``mean (standard deviation)''. All times are in seconds.}
	\label{tblQRall}
\end{table}

The above results can be used to support a conjecture presented by \cite{Imhof}, who stated that a $G$-optimal design for quadratic regression on the (continuous, not discretized) domain $[-1,1]$ should satisfy Theorem 1 of \cite{Imhof} even for $N=5$, although he did not prove this conjecture. That is, such a design should specify one trial to be performed at each of the points $x \in \{-1, -\gamma, 0, \gamma, 1 \}$, where $\gamma \in (0,1)$ is the solution to $-\gamma^4 - 7 \gamma^2 + 4 = 0$. By plotting the corresponding value $\gamma \approx 0.73$ (Figure \ref{fQRall}), we see that the $G$-optimal design indeed satisfies Theorem 1 of \cite{Imhof} (up to the considered discretization). If we add the points $\pm\gamma$ to the set of permissible $x$s, then the $G$-optimal design on the discrete set selects these points and thus exactly corresponds to the $G$-optimal design hypothesized by \cite{Imhof}; see the design labeled ``G-opt. 2'' in Figure \ref{fQRall} (left).

\paragraph{Design constraints} As mentioned in Sections \ref{ssDesignConst} and \ref{ssConstraints}, a beneficial property of the MILP formulation is that it can easily accommodate any constraint that is linear in the design values and in the elements of the covariance matrix $\M^{-1}(\db)$. For instance, the designs in Figure \ref{fQRall} (left) are focused on the regions around $0$, $1$ and $-1$, but one can naturally require a design that is more space-filling. In the present model, we formulate this by introducing the additional constraints
\begin{equation}\label{eConst1}
	\sum_{i=6}^{11} d_i \geq 1, \quad
	\sum_{i=21}^{26} d_i \geq 1.
\end{equation}
These constraints enforce that at least one trial should be performed for $x$ between $-2/3$ and $-1/3$ and also for $x$ between $1/3$ and $2/3$. Such constraints can be included in the linear program, and all the computations can proceed as usual. One exception is that a design $\db_0$ that satisfies the constraints \eqref{eConst1} must be used to construct the relevant covariance bounds. For this purpose, we used a design $\db_0$ that assigns one trial to each of the design points $x \in \{-1, -7/15, 0, 8/15, 1\}$, which satisfies \eqref{eConst1}. The resulting constrained optimal designs obtained via MILP are depicted in Figure \ref{fQRall} (right).

\begin{figure}[h]%
	\centering
	\includegraphics[width=\textwidth]{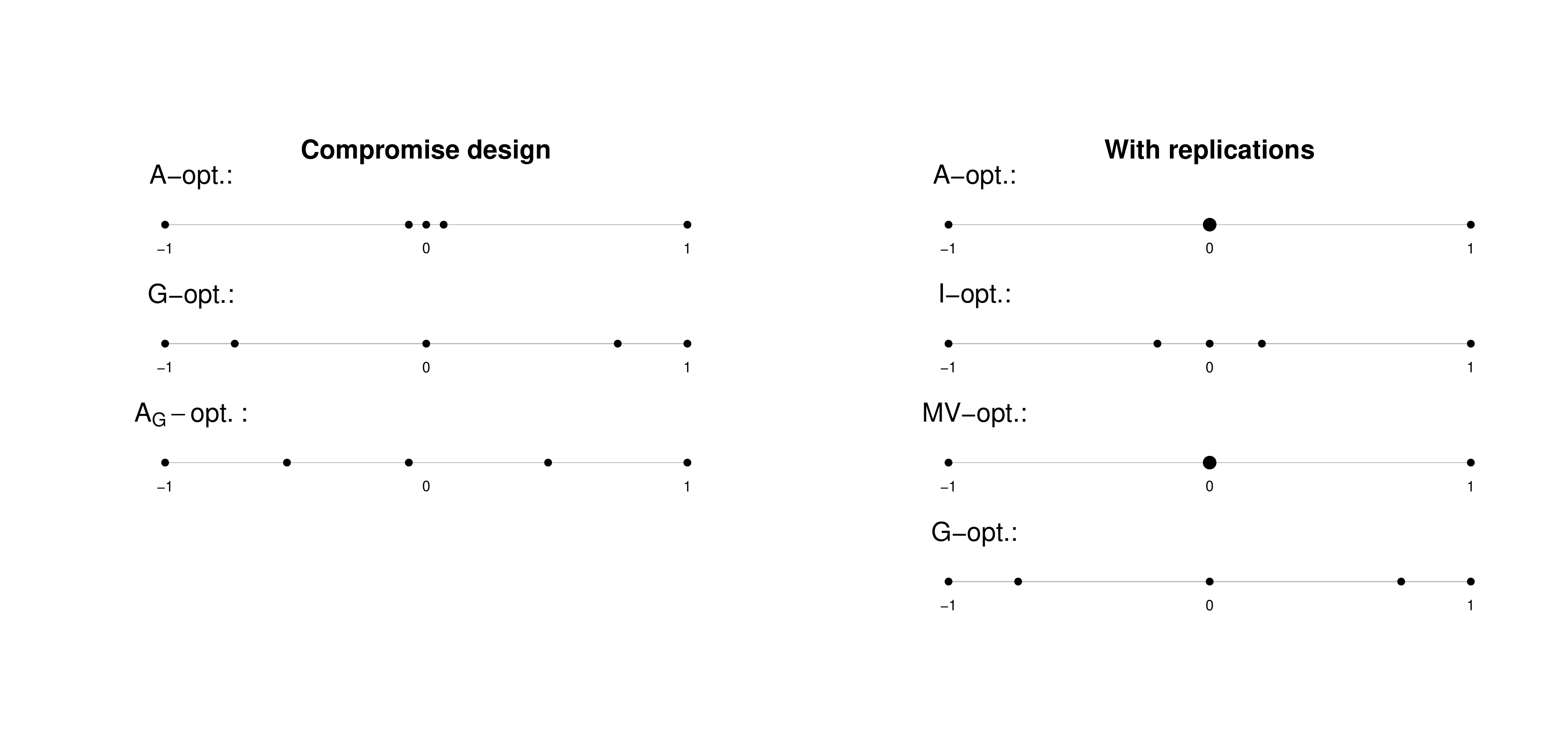}
	\caption{Left: $A$- and $G$-optimal (binary) designs for the quadratic regression model \eqref{eModelQuad} and the design that is $A$-optimal among all the designs $\db$ that satisfy $\Phi_G(\M(\db)) \leq 0.9$. Right: Optimal designs with replications. Smaller dots correspond to $d_i=1$ (one replication), and larger dots correspond to $d_i=3$ (three replications).}\label{fQRsecTogther}
\end{figure}

\paragraph{Covariance constraints} The $A$- and $G$-optimal designs for \eqref{eModelQuad} differ quite significantly, so we might be interested in finding a ``compromise'' design. This can be achieved by, e.g., finding the $A$-optimal design among all the designs that attain at most a specified value of the $G$-optimality criterion. In particular, we have $\Phi_G(\M(\db_A^*)) \approx 1.00$ and $\Phi_G(\M(\db_G^*)) \approx 0.75$, where $\db_A^*$ and $\db_G^*$ are the $A$- and $G$-optimal designs, respectively, and we wish to construct the design $\db_{A_G}^*$ that is $A$-optimal among all designs $\db$ that have $\Phi_G(\M(\db)) \leq 0.9$ (i.e., such designs have a maximum variance of $\f'(x_i)\widehat{\betab}$, $i=1,\ldots,n$, of at most 0.9). The constraints $\Phi_G(\M(\db)) \leq 0.9$ are linear in $\Sigmab = \M^{-1}(\db)$, which means that they can be incorporated into the MILP formulation. To construct the covariance bounds, a design $\db_0$ that satisfies $\Phi_G(\M(\db_0)) \leq 0.9$ must be chosen. As discussed in Section \ref{ssConstraints}, a natural choice is the $G$-optimal design, i.e., $\db_0 = \db_G^*$. The optimal constrained design $\db_{A_G}^*$ obtained via MILP is depicted in Figure \ref{fQRsecTogther} (left), together with $\db_A^*$ and $\db_G^*$. This figure illustrates that $\db_{A_G}^*$ is indeed a compromise  between $\db_A^*$ and $\db_G^*$.

\paragraph{Replicated observations} All the designs discussed above are binary, i.e., without replications. To allow for any number of replications, we can simply repeat each regressor $N$ times and apply the MILP algorithm, as outlined in Section \ref{ssRepeated}. The resulting designs for \eqref{eModelQuad} with $N=5$ are illustrated in Figure \ref{fQRsecTogther} (right), and the computation times are given in Table \ref{tblQRall}. In the $A$- and $MV$-optimal designs, the design point $x_i=0$ is replicated three times, while the $I$- and $G$-optimal designs are still binary even when replications are allowed. The observed computation times demonstrate that allowing for replications can result in a significantly increased running time.

\subsection{Nonlinear regression}
\label{ssNonlinear}

\begin{figure}[h]%
	\centering
	\includegraphics[width=\textwidth]{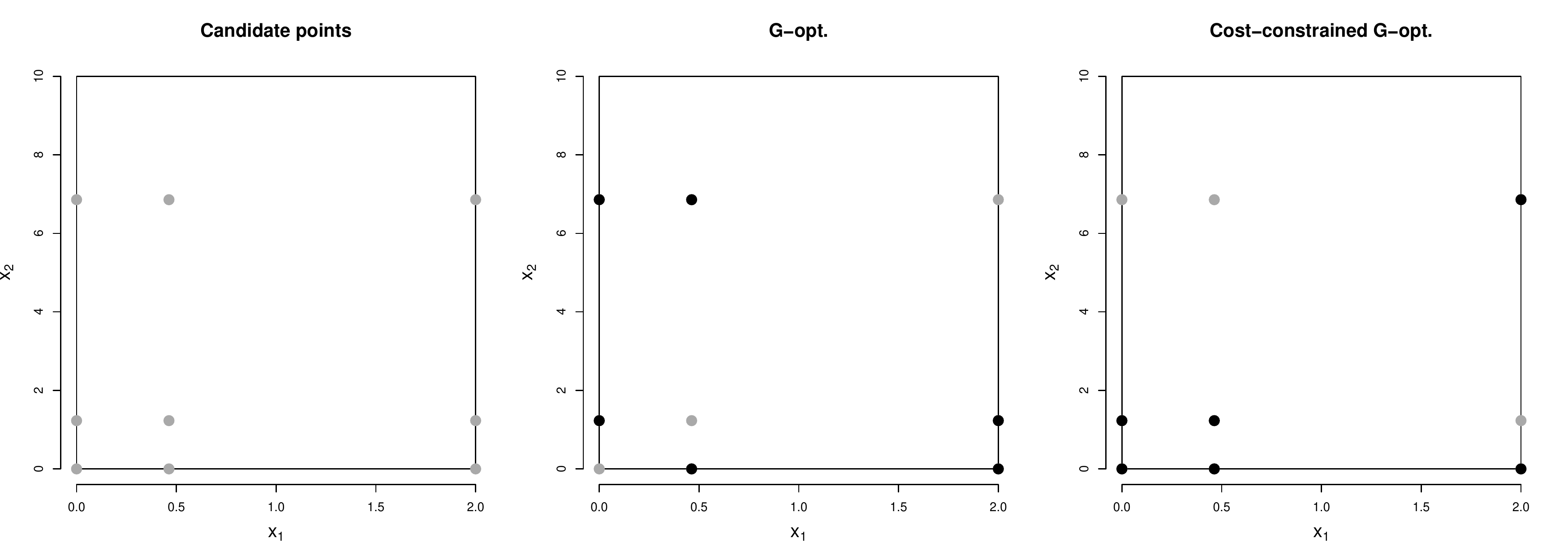}
	\caption{Support points (gray) of the $G$-optimal approximate design for model \eqref{eModPZ} (left), the design (black) that is $G$-optimal among all exact designs supported on these candidate points (middle), and the design (black) that is $G$-optimal among exact designs on these points that satisfy cost constraint \eqref{eCost} with $p_1 = 2$, $p_2 = 1$ and $B=20$ (right).}\label{fNonlinear}
\end{figure}

\paragraph{Nonlinear models, continuous design spaces}
In this section, we demonstrate the usefulness of the MILP approach for nonlinear models and for infinite design spaces. Consider the following model given by \cite{PronzatoZhig} (Problem 10 therein):
\begin{equation}\label{eModPZ}
	y_i = \beta_1 + \beta_2e^{-\beta_3x_{i,1}} + \frac{\beta_4}{\beta_4 - \beta_5}(e^{-\beta_5 x_{i,2}} - e^{-\beta_4 x_{i,2}}) + \varepsilon_i,
\end{equation}
localized at $\betab_0 = (1, 1, 2, 0.7, 0.2)'$, where $\x_i = (x_{i,1}, x_{i,2})' \in [0,2] \times [0,10]$ and $\varepsilon_i$ follows the Gaussian distribution $\mathcal{N}(0, 1)$. Let us seek a $G$-optimal design for $N=6$ trials. Since the design space is continuous, we cannot apply MILP directly. Instead, we first apply the approximate design algorithm GEX proposed by \cite{HarmanEA21} to find the $G$-optimal approximate design (because $G$- and $D$-optimality coincide in the approximate design case, we use the GEX algorithm for $D$-optimality), and we consider the points that support the $G$-optimal approximate design (i.e., $\x_i$ such that $w_i > 0$); see Figure \ref{fNonlinear} (left). We then find the exact design that is $G$-optimal on these nine ``candidate'' support points using our MILP approach; see Figure \ref{fNonlinear} (middle). This design is highly efficient among the designs on the entire design space $[0,2] \times [0,10]$; for instance, it coincides with the design obtained via the KL exchange algorithm for $D$-optimality applied on a densely discretized version of $[0,2] \times [0,10]$. 

Whereas the GEX algorithm took a few seconds and the subsequent MILP computations required less than one second, the KL exchange algorithm required 10 minutes to arrive at the same design. Consequently, for a model with multiple continuous factors, the direct application of an exchange heuristic would be unfeasible; the current relatively simple model was chosen only so that the efficiency of the proposed approach could be easily demonstrated. Note also that the GEX algorithm works by using a dense discretization of the design space (in particular, we considered a grid of more than 20 million points), but any approximate design algorithm for continuous spaces can be used instead, such as particle swarm optimization (cf. \cite{ChenEA22}). In addition, one may employ any nonlinear programming approach for a continuous domain to fine-tune the positions of the support points of the exact design determined by our approach.

\paragraph{Cost constraints} Because the MILP formulation allows for the inclusion of extra constraints, we can consider a cost constraint of the form
\begin{equation}\label{eCost}
	p_1 \sum_{i=1}^9 d_i x_{i,1} + p_2 \sum_{i=1}^9 d_i x_{i,2} \leq B,
\end{equation}
where $p_1$ and $p_2$ represent the costs of factors 1 and 2, respectively; $B$ is the total budget; and $\x_1, \ldots, \x_9$ are the nine candidate points, with $\x_i = (x_{i,1}, x_{i,2})'$. We then use the MILP approach to find the design that is $G$-optimal among all the designs supported on the candidate points satisfying \eqref{eCost} with $p_1 = 2$, $p_2 = 1$ and $B=20$. To construct the covariance bounds for the MILP formulation, we select the design $\db_0$ in which all 6 trials are assigned to the 6 candidate points that satisfy $x_{i,2} < 2$. The optimal cost-constrained design given by the MILP algorithm is depicted in Figure \ref{fNonlinear} (right). 

\subsection{Time complexity}

\begin{figure}[h]%
	\centering
	\includegraphics[width=0.9\textwidth]{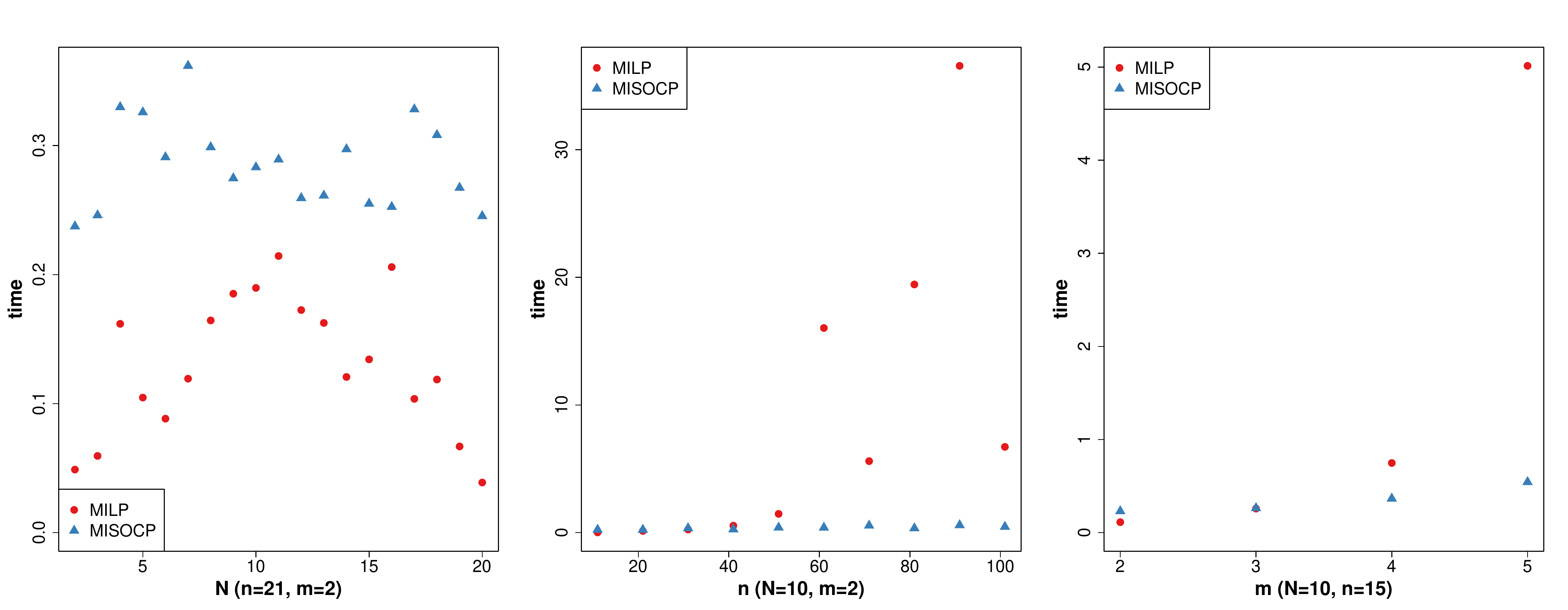}
	\caption{Mean running times of 10 runs of the MILP (red circles) and MISOCP (blue triangles) algorithms for various models of the form \eqref{eModelGaussian}. In each graph, one value ($N$, $n$ or $m$) is varied, while the other two remain fixed---the values of the fixed parameters are specified at the bottom of each graph. All times are in seconds.}\label{fTimeComp}
\end{figure}

To estimate the typical running time of the MILP computation, we consider an (artificial) ``Gaussian'' model 
\begin{equation}\label{eModelGaussian}
	y_i = \f_i' \betab + \varepsilon_i;  \quad \f_i \in \{\f_1, \ldots, \f_n\},
\end{equation}
where the vectors $\f_1, \ldots, \f_n$ are independently generated from the $m$-dimensional standardized Gaussian distribution $\mathcal{N}_m(\0_m, \I_m)$. Figure \ref{fTimeComp} shows the mean running times based on 10 runs of MILP computations of the $A$-optimal designs for various values of $N$, $n$ and $m$. To compare the MILP approach with its most natural competitor, the mean running times of ten runs of the MISOCP approach proposed by \cite{SagnolHarman15}
are also included in this plot. Overall, the MILP algorithm is often faster than the MISOCP algorithm for smaller models, but it tends to fall behind in performance as $n$ and $m$ increase. Note that the two algorithms always returned designs with the same value of the optimality criterion, so we report only their respective computation times.

We also consider three more realistic models: a basic linear regression model
\begin{align}
	y_i &= \beta_1 + \beta_2 x_i + \varepsilon_i, \nonumber \\
	x_i &\in \{-1, -1 + \frac{2}{n-1}, \ldots, 1 - \frac{2}{n-1}, 1\}, \label{eModelLin}
\end{align}
with $n=31$ design points and $m=2$ parameters; the quadratic regression model \eqref{eModelQuad} with $n=21$  and $m=3$; and a factorial main-effects model 
\begin{equation}
    y_i = \beta_1 + \beta_2 x_{i,1} + \ldots + \beta_{m} x_{i,m-1} + \varepsilon_i; \quad
    x_{i, j} \in \{-1,1\} \label{eModelCube}
\end{equation}
with $n=16$ and $m=5$. We compare the MILP and MISOCP computations of the $A$- and $I$-optimal designs for these models for selected numbers of observations $N$; the results are reported in Table \ref{tblMILPvsMISOCP}. 

\begin{table}[h]
	\centering
	\begin{tabular}{c | r r |r r}
		\hline
		\multirow{2}{*}{Model} & \multicolumn{2}{c|}{A-opt.} & \multicolumn{2}{c}{I-opt.} \\
		& $t_{MILP}$ & $t_{MISOCP}$ & $t_{MILP}$ & $t_{MISOCP}$  \\ \hline
		\eqref{eModelLin}, $m=2$, $n=21$, $N=6$  & 0.20 (0.02) & 0.26 (0.02) & 1.05 (0.06) & 0.25 (0.02)
		\\
		\eqref{eModelQuad}, $m=3$, $n=31$, $N=9$  & 24.31 (2.12) & 0.32 (0.07) & 137.69 (7.17) & 1.07 (0.06)
		\\
        \eqref{eModelCube}, $m=5$, $n=16$, $N=7$  &  2.67 (0.30) & 20.49 (0.94) & 2.84 (0.12) & 20.46 (1.17)
        \\
		\hline
	\end{tabular}
	\caption{Running times of the MILP and MISOCP computations of $A$- and $I$-optimal designs for the given models. Each algorithm was run 10 times, and the means and standard deviations of these runs are reported in the form ``mean (standard deviation)''. All times are in seconds.}
	\label{tblMILPvsMISOCP}
\end{table}

The above examples suggest that the MILP approach is often slower than the MISOCP approach for larger $n$ and $m$, but not always. Moreover, even in situations where MILP is currently slower than MISOCP, this may change in the future, e.g., with new developments in MILP solvers (see the Discussion). It should also be noted that the comparisons were performed only for problems with an existing MISOCP formulation. For certain problems that can be solved via our MILP-based approach, such as $MV$-optimality problems and problems with covariance or design constraints,
we are unaware of available MISOCP implementations; hence, corresponding comparisons were not feasible.

\section{Discussion}
\label{sDiscussion}

We have shown that a large class of optimal exact design problems can be formulated as mixed-integer linear programs. Although our implementation of the MILP computation of exact designs is generally comparable to the existing MISOCP approach, there are several important advantages of the MILP approach:
\begin{enumerate}
	\item The MILP formulation is important from a theoretical perspective---it shows that many optimal design problems can be solved using solvers that are even more specialized than MISOCP solvers, continuing the ``increasing specialization'' trend of mixed-integer programming $\to$ mixed-integer second-order cone programming $\to$ mixed-integer linear programming.
	\item We provide an MILP formulation for criteria and design problems for which, to our knowledge, no MISOCP formulation has been published---for example, $MV$-optimality and optimal designs under constraints on the (co)variances of the least-squares estimator.
	\item MILP solvers are more readily available than MISOCP solvers. For example, there are freely available solvers for MILP (e.g., the package \texttt{lpSolve}) in \texttt{R}, whereas the MISOCP solvers in \texttt{R} all seem to be dependent on commercial software (such as \texttt{gurobi}), although currently with free licenses for academics.
	\item It is possible that the MILP algorithm may become a faster means of computing optimal designs than the MISOCP algorithm, with either further developments in MILP solvers, the use of different solvers, the application of particular tricks such as ``symmetry breaking'', or perhaps a different MILP formulation. Moreover, in this study, we used only the default settings, as our aim was not to fine-tune the solver for particular problems but rather to examine its expected behavior---as such, it may be possible to increase the speed of the MILP computations by using customized solver settings.
\end{enumerate}
It is therefore important to know that the considered optimal design problems can be expressed in MILP form at all so that, starting from this foundation, the formulation and implementation can be further improved.

It is also worth noting that methods with a guarantee of optimality (such as MILP and MISOCP) are generally limited to either small or medium-sized problems due to their time complexity, unlike various exchange-type heuristics. However, in classical statistical applications, the problem size is often not too large (e.g., the number $m$ of variables is usually less than $10$). In practice, therefore, we would recommend first attempting to use an ``exact'' method such as MILP or MISOCP for a given problem and applying a heuristic method only if the exact method takes too long to find a solution or if the problem is clearly too large. Moreover, the MILP and MISOCP formulations can accommodate additional constraints that are beyond the capabilities of available heuristic methods.

We also emphasize that the MILP approach, similarly to other algorithms for discrete design spaces, can also be used for solving relevant subproblems even for continuous design spaces, as demonstrated in Section \ref{ssNonlinear}.

Finally, we note that the provided MILP formulation opens up interesting research avenues: e.g., the analytical and computational construction of better covariance bounds, and the construction of designs $\db_0$ for arbitrary design or covariance constraints.

\section*{Declarations}

Declarations of interest: none.

\section*{Acknowledgments}

This work was supported by the Slovak Scientific Grant Agency [grant VEGA 1/0362/22].

\appendix

\section{MILP formulation}\label{sAppA}

Here, we give the exact mathematical form of the MILP formulation \eqref{ePhiWLin} that can be used as an input to MILP solvers. Note that we use the $\vv$ operator (the vectorization of a matrix obtained by stacking its columns), but due to symmetry, the more efficient formulation with $\vech$ (the vector-half operator for symmetric matrices) for both $\Sigmab$ and $\z$ can be used instead. However, the $\vech$ version seems too complicated to describe concisely here. 

For simplicity, we first consider $A$-optimality, i.e., $\Phi_\Bf(\M) = \tr(\M^{-1})$. For $A$-optimality, formulation \eqref{ePhioptLin} is still linear and simpler than \eqref{ePhiWLin}, so we actually describe \eqref{ePhioptLin} here. Let $\cb = \vv(\Sigmab) \in \R^{m^2}$, $\M_i = \f_i\f_i'$, $\Z_i = (z_{ijk})_{j,k=1}^m \in \R^{m \times m}$, $\z_i = \vv(\Z_i) \in \R^{m^2}$ (for $i=\{1{:}n\}$), $\z = (\z_1', \ldots, \z_n')'$, and $\x = (\z', \db', \cb')'$. Overall, we have the vector of variables $\x = (\z', \db', \cb')'$, where $\z \in \R^{nm^2}$, $\db \in \{0,1\}^n$ and $\cb \in \R^{m^2}$. Below, we express the constraints and the objective function in \eqref{ePhioptLin} using $\x$.

1. $\M \Sigmab = \I_m$: The left-hand side can be expressed as
$$
\M \Sigmab = \sum_{i=1}^n d_i \M_i \Sigmab = \sum_{i=1}^n \M_i \Z_i.
$$
By applying the $\vv$ operator, we obtain
$
\sum_i (\I_m \otimes \M_i) \vv(\Z_i) = \vv(\I_m)
$
because $\vv(\A\B) = (\I \otimes \A) \vv(\B)$. It follows that $\M \Sigmab = \I_m$ can be expressed as
$$
\begin{pmatrix}
	\I_m \otimes \M_1 & \ldots & \I_m \otimes \M_n & \0_{m^2 \times n} & \0_{m^2 \times m^2}
\end{pmatrix} \x = \vv(\I_m)
$$
(the matrices of zeros correspond to $\db$ and $\cb$ in $\x$).

2. $z_{ijk} - d_i L_{jk} \geq 0$: For each $i$, this can be expressed as $\vv(\Z_i) - d_i \vv(\Lb) \geq \0_{m^2}$, where $\Lb = (L_{jk})_{j,k=1}^m \in \R^{m \times m}$. It follows that $z_{ijk} - d_i L_{jk} \geq 0$ for all $i \in \{1{:}n\}$ is equivalent to 
$$\begin{pmatrix}
	\I_{nm^2} & -\I_n \otimes \vv(\Lb) & \0_{nm^2 \times m^2}
\end{pmatrix}\x \geq \0_{nm^2}.$$

3. The other constraints, \eqref{eMcCormick2}-\eqref{eMcCormick4}, can be similarly expressed as
$$\begin{pmatrix}
	\I_{nm^2} & -\I_n \otimes \vv(\U) & -\1_n \otimes \I_{m^2}
\end{pmatrix}\x \geq -\1_n \otimes \vv(\U),$$
$$\begin{pmatrix}
	\I_{nm^2} & -\I_n \otimes \vv(\U) & \0_{nm^2 \times m^2}
\end{pmatrix}\x \leq \0_{nm^2},$$
$$\begin{pmatrix}
	\I_{nm^2} & -\I_n \otimes \vv(\Lb) & -\1_n \otimes \I_{m^2}
\end{pmatrix}\x \leq -\1_n \otimes \vv(\Lb).$$

4. $\1_n'\db = N$: This is clearly equivalent to $(\0_{nm^2}', \1_n', \0_{m^2}') \x = N$.

5. The objective function for $A$-optimality is $\tr(\Sigmab) = \ab'\x$, where $\ab' = (\0_{nm^2}', \0_n', \vv(\I_m)')$.
\bigskip

Constraints 1-4 and objective function 5 above form the MILP formulation for $A$-optimality. For a general $\Phi_\Bf$ criterion, we have $\varphi \geq \tr(\B_\ell' \Sigmab \B_\ell)$ for $\ell=\{1{:}K\}$ in the more general formulation \eqref{ePhiWLin}. The vector $\x$ is then redefined as $\x = (\z', \db', \cb', \varphi)'$, and the objective function in item 5 above becomes $\ab'\x$, where $\ab' = (\0_{nm^2}', \0_n', \0_{m^2}', 1)$, and a column of zeros representing $\varphi$ is added to each left-hand-side matrix in constraints 1-4. 

6. An additional set of constraints, representing $\varphi \geq \tr(\B_\ell' \Sigmab \B_\ell)$, also needs to be included. These constraints can be expressed by setting $\G_\ell = \B_\ell\B_\ell'$ ($\ell=\{1{:}K\}$):
$$
\varphi \geq \tr(\B_\ell' \Sigmab \B_\ell) = \tr(\Sigmab \G_\ell) = \vv(\G_\ell)' \vv(\Sigmab),
$$
which means that the constraints are
$$
\begin{pmatrix}
	\0_{nm^2}' & \0_n' & -\vv(\G_1)' & 1 \\
	\vdots & \vdots & \vdots & \vdots \\
	\0_{nm^2}' & \0_n' & -\vv(\G_K)' & 1
\end{pmatrix} \x \geq \0_{K}.
$$

\section{Proofs}\label{sAppB}

Here, we provide the nontrivial proofs of the lemmas and theorems presented in the paper.

\begin{proof}[Lemma \ref{lemma1}]
	Let $\Sigmab \in \Sbb^m_{+}$ be such that $\tr(\B_\ell'\Sigmab \B_\ell) \leq \alpha$ for all $\ell \in \{1{:}K\}$. Let $\w=(w_1,\ldots,w_K)' \in \mathbb{R}^K$ be a vector with nonnegative components summing to $1$, and let $\N:=\N(\w)=\sum_{\ell=1}^K w_\ell \B_\ell \B'_\ell$. First, since $\max_\ell \tr(\B'_\ell \Sigmab \B_\ell) \leq \alpha$, we have
	\begin{align}
		\tr(\N^{1/2}\Sigmab\N^{1/2})&=\tr(\N\Sigmab)& \nonumber \\
		&=\sum_\ell w_\ell \tr(\B'_\ell \Sigmab \B_\ell) \leq \alpha. \label{eq:T10}
	\end{align}
	Consider an $m \times r$ matrix $\X$ such that $\C(\X) \subseteq \C(\N)$. We have
	\begin{align}
		&\tr(\X' \Sigmab \X)&  \\
		&= \tr(\X'(\N^{1/2})^+\N^{1/2}\Sigmab\N^{1/2}(\N^{1/2})^+\X)& \label{eq:pT11} \\
		&= \tr([\N^{1/2}\Sigmab\N^{1/2}][(\N^{1/2})^+\X\X'(\N^{1/2})^+])& \label{eq:pT12} \\
		&\leq \tr(\N^{1/2}\Sigmab\N^{1/2})\lambda_{\max}((\N^{1/2})^+\X\X'(\N^{1/2})^+)& \label{eq:pT13}\\
		&\leq \alpha \lambda_{\max}(\X'\N^+\X),& \label{eq:pT14}
	\end{align}
	where \eqref{eq:pT11} follows from $\X=\N^{1/2}(\N^{1/2})^+\X$, \eqref{eq:pT12} is a basic property of the trace, \eqref{eq:pT13} follows from $\tr(\Q\Hb)\leq \tr(\Q)\lambda_{\max}(\Hb)$ for any $\Q,\Hb \in \Sbb^m_{+}$, and \eqref{eq:pT14} is valid because of \eqref{eq:T10}, $\lambda_{\max}(\Y\Y')=\lambda_{\max}(\Y'\Y)$ for any matrix $\Y$, and $(\N^{1/2})^+(\N^{1/2})^+=\N^+$ (e.g., Example 7.54 (c) in \cite{Seber}).
\end{proof}

\begin{proof}[Lemma \ref{lem:C}]
	We need only to prove that $\e_j \in \C(\N(\w^{(j+)}))$ for any $j \in \{1{:}m\}$; all other claims of the lemma are then trivial. We fix $j \in \{1{:}m\}$ and adopt the notations $\w:=(w_1,\ldots,w_K)':=\w^{(j+)}$, $\h:=(\h'_1,\ldots,\h'_K)':=\B^+\e_j$, where $\h_1 \in \R^{s_1},\ldots,\h_K \in \R^{s_K}$, and $\tilde{\B}=(\sqrt{w_1}\B_1,\ldots,\sqrt{w_K}\B_K)$. Because $\e_j \in \C(\B\B')=\C(\B)$, we have
	\begin{align*}
		\e_j &= \B\B^+\e_j = \B\h = \sum_\ell \B_\ell \h_\ell = \sum_{\ell; \h_\ell \neq \0_{s_\ell}} \B_\ell \h_\ell \\
		&= \sum_{\ell; w_\ell>0} 
		\sqrt{w_\ell} \B_\ell (\h_\ell/\sqrt{w_\ell}) = \tilde{\B}\tilde{\h},
	\end{align*}
	where $\tilde{\h}=(\h'_1/\sqrt{w_1},\ldots,\h'_K/\sqrt{w_K})'$ and $\0/0:=\0$. That is, $\e_j \in \C(\tilde{\B})$. However, $\C(\tilde{\B}) = \C(\tilde{\B}\tilde{\B}') = \C(\N(\w))$. 
\end{proof}

\bibliographystyle{elsarticle-harv} 
\bibliography{rosa.bib}

\end{document}